\documentclass[12pt]{article}
\usepackage{amsfonts,psfig}
\usepackage{epsfig}
\usepackage{graphics}
\begin{document} 
\vspace*{3cm}
\begin{center}
{\bf NLO CHPT corrections to the transverse muon polarization in $K_{l2\gamma}$ decay.}
\end{center}

\vspace*{1cm}
\begin{center}
V.V. Braguta, A.E. Chalov, A.A. Likhoded$^{\dagger}$,
\end{center}

\vspace*{1cm}
\begin{center}
$^{\dagger}$ {\it Institute for high energy physics, Protvino} \\
\end{center}

\vspace*{2cm}
\underline{\bf Abstract}

\vspace*{0.5cm}
\noindent
The calculation of the transverse muon polarization in $K^+ \to \mu^+ \nu \gamma$ decay
due to the final state interaction up to $O(p^6)$ CHPT level has been carried out. 
It is shown that $O(p^6)$ corrections to the observable is much less than $O(p^4)$ contribution.
\newpage
\section{Introduction}

\noindent
Despite considerable success of the Standard Model(SM) in predictions of many experiments 
the question about the mechanism of $CP$-violation still remains unresolved. 
In the framework of  SM $CP$-violation results from the nonzero phase in CKM 
matrix, but there are many models with different mechanisms of $CP$-violation. 
In order to reveal the mechanism of this phenomena it is interesting to 
measure physical observables that are very sensitive to  $CP$-violation 
beyond SM. 

In our paper we will consider transverse muon polarization in $K^+\to \mu^+\nu\gamma$ decay.
$CP$-violating contribution to this observable in the framework of SM is negligible. 
At the same time some mechanisms of $CP$-violation different from that in SM could 
give considerable contribution to the value of the transverse polarization\cite{bezrukov}. 
That is why it is interesting to measure the transverse muon polarization at experiment. 

In addition to the negligible $CP$-violating contribution the nonzero value of the transverse 
muon polarization in the framework of SM can originate from final state interaction effect
what can be a real obstacle in a new physics searches. The effect of final state interaction 
was considered in papers  \cite{braguta,rogalyov} up to $O(p^4)$ in CHPT. 
It was shown that the average value of the transverse polarization is $P_T= 5.6 \times 10^{-4}$. 
Unfortunately, current experiments are unable to reach the accuracy comparable to the predictions
in the framework of SM( for instance the accuracy of KEK-E246 experiment is  $\sim 10^{-2}$\cite{kudenko}) 
But, future experiments will be able measure the transverse muon polarization in $K^+\to \mu^+\nu\gamma$
decay with sufficient accuracy. For instance, it is planning to get $\sim 10^{10}$ events
of this decay for one year at the experiment JHL\cite{furusaka}. Such statistic 
allows one to measure the transverse muon polarization with the accuracy $\sim 10^{-4} - 10^{-5}$. 

Obviously high accuracy in measurement of the transverse muon polarization imposes 
additional requirements to the theoretical prediction for the value of this observable 
in the framework of SM due to the final state interaction effect. 
The aim of our paper is to consider the effect of final state interaction up 
to $O(p^6)$ in CHPT.  

Our paper is organized as follows: next section is devoted to calculation of 
the transverse muon polarization up to $O(p^6)$ in CHPT. In last section we 
present the discussion of the results obtained in our paper. 

\section{Transverse muon polarization in $K^+ \to \mu^+ \nu \gamma$ decay}

\noindent
The $K^+ \rightarrow \mu^+ \nu \gamma$ decay at tree level of SM
is described by the diagrams shown in  Fig.~1. The diagrams in Fig. 1b and 1c
correspond to the muon and kaon bremsstrahlung, while the diagram in Fig. 1a 
corresponds to the structure radiation.
This decay amplitude can be written as follows
\begin{equation}
M=ie \frac{G_F}{\sqrt{2}}V^{*}_{us}\varepsilon^{*}_{\mu}\left(f_K m_{\mu}
\overline u(p_{\nu})(1+\gamma_{5}) \biggl( \frac{p_{K}^{\mu}}{(p_K q)}
-\frac{(p_{\mu})^{\mu}}{(p_{\mu} q)}-
\frac{\hat{q} \gamma^{\mu}}{2( p_{\mu} q)} \biggr)v(p_{\mu})-G^{\mu \nu}
l_{\nu}
\right) \; ,
\end{equation}
where 
\begin{eqnarray}
l_{\mu}=\overline u(p_{\nu}) (1+\gamma_{5}) \gamma_{\mu}
v(p_{\mu}) \; ,\nonumber \\
G^{\mu \nu}= i F_v ~\varepsilon^{ \mu \nu \alpha \beta } q_{\alpha}
(p_K)_{\beta} -
F_a ~ ( g^{\mu \nu} (p_K q)-p_K^{\mu} q^{\nu} )\;, 
\end{eqnarray}
$G_F$ is the Fermi constant, $V_{us}$ is the corresponding 
CKM matrix element, $f_K$ is the $K$-meson leptonic constant,
$p_K$, $p_\mu$, $p_\nu$, $q$ are the kaon, muon, neutrino, and photon four-momenta,
correspondingly, and $\varepsilon_{\mu}$ is the photon polarization vector. 
$F_v$ and $F_a$ are the kaon vector and axial formfactors. Here it is worth 
to mention that formfactors $F_v$ and $F_a$ do not depend on the momentum 
of virtual $W$-boson and are equal to 
\begin{eqnarray}
F_v^0=\frac{0.095}{m_K},\;\; F_a^0=-\frac{0.043}{m_K}\;.
\label{ff}
\end{eqnarray}
But if one considers these formfactors at $O(p^6)$ level there appears the dependence on 
the momentum of virtual $W$-boson. The calculation of this dependence 
was carried out in paper \cite{corr}. Unfortunately it is not possible to 
use this result since there are many unknown constants. That is why 
the model of vector meson dominance will be used in our calculation:
\begin{eqnarray}
\nonumber
F_v(W^2)= \frac {F_v^0} {1-W^2/M_{K^*}^2} \simeq F_v^0 (1+ \frac {W^2} {M_{K^*}^2} ) \\ 
F_a(W^2)= \frac {F_a^0} {1-W^2/M_{K_1}^2} \simeq F_a^0 (1+ \frac {W^2} {M_{K_1}^2} ) ,
\end{eqnarray}
where the following designations are used: $M_{K^*}, M_{K_1}$- the masses of $K^*$ and $K_1$ mesons, $W^2=(p_K-q)^2$.
In Eq. (2) we use the following
definition of Levi-Civita tensor: $\epsilon^{0 1 2 3} = +1$.

The partial width of the $K^+ \to \mu^+ \nu \gamma$ decay in the  
$K$-meson rest frame can be expressed as 
\begin{equation}
d \Gamma = \frac {\sum |M|^2} { 2 m_K } (2 \pi)^4 \delta ( p_K - p_\mu -q
-p_\nu)
\frac {d^3 q} {(2 \pi)^3 2 E_q } \frac {d^3 p_\mu} {(2 \pi)^3 2 E_\mu }
\frac {d^3 p_\nu} {(2 \pi)^3 2 E_\nu }\;,
\end{equation}
where summation over muon and photon spin states is performed.

Introducing the unit vector along the muon spin direction in muon rest frame,
$\bf \vec s $, where ${\bf \vec e}_i~(i=L,\: N,\: T)$ are the unit vectors
along the longitudinal, normal and transverse components of muon polarization, 
one can write down the matrix element squared for  the transition
into the particular muon polarization state in the following form:
\begin{equation}
|M|^2=\rho_{0}[1+(P_L {\bf \vec e}_L+P_N {\bf \vec e}_N+P_T {\bf \vec
e}_T)\cdot \bf \vec s]\;,
\end{equation}
where $\rho_{0}$ is the Dalitz plot probability density averaged over polarization states.
The  unit vectors ${\bf \vec e}_i$ can be expressed in terms of the
three-momenta of final particles:
\begin{equation}
{\bf \vec e}_L=\frac{\vec p_{\mu}}{|\vec p_{\mu}|}~~~ 
{\bf \vec e}_N=\frac{\vec p_{\mu}\times(\vec q \times \vec p_{\mu})}
{|\vec p_{\mu}\times(\vec q \times \vec p_{\mu})|}~~~
{\bf \vec e}_T=\frac{\vec q \times \vec p_{\mu}}{|\vec q \times \vec
p_{\mu}|}\: . 
\end{equation}
With such definition of ${\bf \vec e}_i$ vectors, $P_T, P_L$, and $P_N$ denote
transverse, longitudinal, and normal components of the muon polarization,
correspondingly. It is convenient to use the following variables
\begin{equation}
x=\frac{2 E_\gamma}{m_K}\;,~~~y=\frac{2 E_\mu}{m_K}\;,
~~~\lambda=\frac{x+y-1-r_\mu}{x}\;,~~~r_{\mu}=\frac {m_{\mu}^2}{m_K^2}\;,
\end{equation}
where $E_\gamma$ and $E_\mu$ are the photon and muon energies in the kaon
rest frame. 

The Dalitz plot probability density, as a function of the  $x$ and $y$
variables, has the form:
\begin{eqnarray}
\rho_{0}=\frac{1}{2}e^2 G_F^2 |V_{us}|^2 &\cdot & \biggl( 
\frac {4 m_{\mu}^2 |f_K|^2 } {\lambda x^2} (1-\lambda) \Bigl (x^2+2 (1-r_{\mu})
(1-x-\frac {r_{\mu}} {\lambda} )\Bigr)	\nonumber\\
&+& m_K^6 x^2 (|F_a|^2+|F_v|^2) (y-2 \lambda y-\lambda x +2
\lambda^2)\nonumber\\
&+& 4  ~\hbox{Re} (f_K F_v^*) ~m_K^4 r_{\mu} \frac x {\lambda} (\lambda -1)
\hspace{0.3in} \nonumber\\
&+&4 ~\hbox{Re} (f_K F_a^*) ~m_K^4  r_{\mu} (-2 y+x+2 \frac {r_{\mu}} {\lambda}
-
\frac x {\lambda} +2 \lambda )\hspace{0.3in} \nonumber\\
&+& 2 ~\hbox{Re} (F_a F_v^*) ~m_K^6 x^2 (y-2 \lambda +x \lambda) \biggr)\;.
\end{eqnarray}
Calculating the muon transverse polarization $P_T$ we follow the original paper [7]
and assume that the decay amplitude  is $CP$-invariant,
and formfactors $ f_K $, $ F_v$, and $ F_a$ are real.
In this case the tree level muon polarization $P_T=0$.
When one-loop contributions are incorporated, the nonvanishing muon
transverse polarization can arise due to the interference of tree-level 
diagrams and imaginary parts of one-loop diagrams, induced by the electromagnetic 
final state interaction.

To calculate the imaginary parts of formfactors one can use the
$S $-matrix unitarity:
\begin{equation}
S^+ S=1
\end{equation}
and, using $ S=1+i T$, one gets
\begin{equation}
T_{f i}-T_{i f}^*=i \sum_n T^*_{n f} T_{n i}\;, 
\end{equation}
where $i,\: f, \: n$ indices correspond to the initial, final, and intermediate
states of the particle system.
Further, using the  $T$-invariance of the matrix element one has
\begin{eqnarray}
\mbox{Im} T_{f i}=\frac 1 2 \sum_n T^*_{n f} T_{n i}\;, 
\label{im}
\end{eqnarray}
In the framework of SM one-loop diagrams that contribute to the muon transverse polarization in
the $K^+ \to \mu^+ \nu \gamma$ decay, are shown in Fig. 2.
The calculation of the transverse muon polarization up to $O(p^4)$ has already 
been carried out in papers \cite{braguta, rogalyov}. The calculation of $O(p^6)$
corrections to the observable will be done taking into the account the dependence 
of the formfactors $F_v, F_a$ on the momentum of virtual $W$-boson. In other 
words one should carry out similar calculation as was done in papers  \cite{braguta}
with the substitution :
\begin{eqnarray}
\nonumber
F_v^0 \to F_v^0 (1+ \frac {W^2} {M_{K^*}^2} ) \\ 
F_a^0 \to F_a^0 (1+ \frac {W^2} {M_{K_1}^2} ).
\label{zamena}
\end{eqnarray}
The contribution of $O(p^6)$ corrections to the transverse muon polarization can be separated into two parts. 
The calculation of the first part is very simple: one should make substitution (\ref{zamena})
in the expressions obtained in paper \cite{braguta}. 
Let us proceed with the calculation of the second part.

As it was already noted the diagrams that give nonzero contribution to 
the transverse muon polarization are presented in Fig. 2. The diagrams presented in 
Fig. 2 a,b,e,f do not lead to the additional $O(p^6)$ contribution due to the replacement
(\ref{zamena}). The diagram in Fig. 2g will give additional contribution to 
the the transverse muon polarization if one regards $O(p^6)$ corrections. But $O(p^4)$ contribution
of this diagram is strongly suppressed because of suppression of the intermediate state 
phase space. We will suppose that the same suppression holds for $O(p^6)$ contribution
of this diagram to the whole answer of $O(p^6)$ corrections and we will not regard 
this diagram in further analysis. As the result the only diagrams presented  in Fig. 2 c,d 
will be considered in our paper. Using expression (\ref{im}) it is possible to write imaginary 
parts of these diagrams. The contribution of diagram in Fig. 2c has the form
\begin{eqnarray}
~\hbox{Im}M_1=\frac{i e \alpha}{2 \pi}\frac{G_F}{\sqrt{2}}V_{us}^*
\overline u(p_{\nu})(1+\gamma_{5}) \int \frac{d^3 k_{\gamma}}{2 \omega_{\gamma}}
\frac{d^3 k_{\mu}}{2 \omega_{\mu}}\delta(k_{\gamma}+k_{\mu}-P)R_{\mu}\times \nonumber\\
(\hat k_{\mu}-m_{\mu})\gamma^{\mu}\frac{\hat q+ \hat p_{\mu}-m_{\mu}}{(q+p_{\mu})^2-m_{\mu}^2}
\gamma^{\delta} \varepsilon^*_{\delta} v(p_{\mu})	\hspace{1.in}
\label{int1}
\end{eqnarray}
The contribution of diagram in Fig. 2c has the form
\begin{eqnarray}
~\hbox{Im}M_2=\frac{i e \alpha}{2 \pi}\frac{G_F}{\sqrt{2}}V_{us}^*
\overline u(p_{\nu})(1+\gamma_{5}) \int \frac{d^3 k_{\gamma}}{2 \omega_{\gamma}}
\frac{d^3 k_{\mu}}{2 \omega_{\mu}}\delta(k_{\gamma}+k_{\mu}-P)R_{\mu}\times \nonumber\\
(\hat k_{\mu}-m_{\mu})\gamma^{\delta} \varepsilon^*_{\delta} 
\frac{ \hat k_{\mu}-\hat q-m_{\mu}}{(k_{\mu}-q)^2-m_{\mu}^2}
\gamma^{\mu} v(p_{\mu}), 
\label{int2}
\end{eqnarray}
where we have used the following designation
\begin{equation}
R_{\mu}= -i F_v^0 \biggl ( \frac {W^2} {M_{K^*}^2} \biggr ) \varepsilon_{\mu \nu \alpha \beta} 
(k_{\gamma})^{\alpha} (p_K)^{\beta} \gamma^{\nu}+
F_a^0 \biggl ( \frac {W^2} {M_{K_1}^2} \biggr ) (\gamma_{\mu}  (p_K k_{\gamma})-(p_K)_{\mu} \hat k_{\gamma})\:.
\end{equation}
The integrals that appear in (\ref{int1}), (\ref{int2}) are given in Appendix.

The expression for the amplitude including Im$M_1$ +Im$M_2 $ can be written as:
\begin{eqnarray}
M=ie \frac{G_F}{\sqrt{2}}V^{*}_{us}\varepsilon^{*}_{\mu}\Biggl(\tilde f_K m_{\mu}
\overline u(p_{\nu})(1+\gamma_{5}) \biggl( \frac{p_{K}^{\mu}}{(p_K q)}
-\frac{(p_{\mu})^{\mu}}{(p_{\mu} q)}
\biggr)v(p_{\mu})+ \nonumber\\
\tilde F_n \overline u(p_{\nu})(1+\gamma_{5}) \hat{q} \gamma^{\mu} v(p_{\mu})
-\tilde G^{\mu \nu} l_{\nu}\Biggr),\hspace{1.in}
\end{eqnarray}
где 
\begin{equation}
\tilde G^{\mu \nu}= i \tilde F_v ~\varepsilon^{ \mu \nu \alpha \beta } q_{\alpha} (p_K)_{\beta} -
\tilde F_a ~ ( g^{\mu \nu} (p_K q)-p_K^{\mu} q^{\nu} ) \;.
\end{equation}
The $\tilde f_K ,\;  \tilde F_v ,\; \tilde F_a$, and 
$\tilde F_n$ formfactors include one-loop contributions 
from the diagrams shown in Figs.~2.  
The choice of the formfactors is determined by the matrix element expansion
into set of gauge-invariant structures.

Since we are interested in the contributions of 
imaginary parts of one-loop diagrams only (since they lead to a
nonvanishing value of the transverse polarization), 
we neglect the real parts of these diagrams and assume 
that $\mbox{Re} \tilde f_K ,\; \mbox{Re} \tilde F_v ,\; \mbox{Re}
\tilde F_a $ coincide with their tree-level values,
$f_K ,\; F_v ,\;  F_a $, correspondingly, and  
$ \mbox{Re} \tilde F_n= - f_K m_\mu /2(p_\mu q)$.
Using the expresions of the integrals given in Appendix it is not difficult 
to calculate $O(p^6)$ contribution. But the explicit result
is rather cumbersome and we will not present it in our paper.

The muon transverse polarization can be written as
\begin{equation}
P_T=\frac {\rho_T} {\rho_0},
\end{equation}
где
\begin{eqnarray}
\rho_T=-2 m_K^3 e^2 G_F^2 |V_{us}|^2 x \sqrt{\lambda y - \lambda^2 - r_{\mu}}
\biggl(
m_{\mu} ~\hbox{Im}(\tilde f_K \tilde F_a^*)(1-\frac{2}{x}+
\frac{y}{\lambda x})+ \nonumber\\ 	
m_{\mu} ~\hbox{Im}(\tilde f_K \tilde F_v^*)(\frac{y}{\lambda x}-1-2 \frac{r_{\mu}}{\lambda x})+
2 \frac{r_{\mu}}{\lambda x} ~\hbox{Im}(\tilde f_K \tilde F_n^*)(1-\lambda)+ \hspace {0.4in}\nonumber\\  
m_K^2 x ~\hbox{Im}(\tilde F_n \tilde F_a^*)(\lambda-1)+ m_K^2 x ~\hbox{Im}
(\tilde F_n \tilde F_v^*)(\lambda-1)\biggr)		
\hspace{0.6in}
\end{eqnarray}

\section{Discussion}

\noindent

For the numerical calculations we use the following formfactor values:
$$ 
f_K=0.16 \mbox{ GeV},\;\; F_v^0=\frac{0.095}{m_K},\;\; F_a^0=-\frac{0.043}{m_K}\;.
$$
The $ f_K $ formfactor is determined from experimental data on kaon decays \cite{rewiev},
and $F_v^0, F_a^0 $ ones are calculated at the one loop-level in the Chiral
Perturbation Theory \cite{bijnens}.
 
Averaged value of the transverse muon polarization integrated over the physical region with the 
cut on photon energy $E_\gamma > 20 $~MeV is equal to
\begin{equation}
\langle P^{SM}_T \rangle = 3 \times 10^{-6}\;. 
\label{resul1}
\end{equation}

The level curves of $O(p^6)$ corrections to the transverse muon polarization are 
shown in Fig. 3. From this figure one sees that the characteristic value 
of $O(p^6)$ corrections to the observable under consideration is about $10^{-5}-10^{-6}$. 
To compare $O(p^6)$ corrections with the  transverse muon polarization with $O(p^4)$ 
result obtained in paper \cite{braguta} the three-dimensional distribution and level curves of the muon transverse polarization
are shown in Fig. 4 and Fig. 5.

\section*{Acknowledgements}

\noindent
The authors thank Dr. Likhoded A.K. for fruitful discussion and valuable remarks. 
This work was in part supported by the Russian Foundation for Basic Research,
grant 04-02-17530, Russian Education Ministry, grant 
E02-31-96, CRDF MO-011-0, Scientific School, grant SS-1303.2003.2 and Dynasty foundation.

\newpage
\section*{Appendix}			  

\noindent
For the the integrals which contribute to (\ref{int1}) and (\ref{int2}),  we use the following
notations:
$$
P=p_\mu+q
$$
$$
d \rho =\frac {d^3 k_{\gamma}} {2 \omega_{\gamma}}
\frac {d^3 k_{\mu}} {2 \omega_{\mu}}\delta(k_\gamma+k_{\mu}-P) 
$$
It should be noted that most of integrals in expressions (\ref{int1}) and (\ref{int2})
were calculated in paper \cite{braguta}. There are only two integrals to be calculated.
The first integral has the form:
\begin{eqnarray}
\int d \rho k_{\gamma}^{\alpha} k_{\gamma}^{\beta} k_{\gamma}^{\delta} = a_{14} 
( g^{\alpha \beta} P^{\delta} + g^{\alpha \delta} P^{\beta} + g^{\beta \delta} P^{\alpha})
+b_{14} P^{\alpha} P^{\beta} P^{\delta}\;. \nonumber 
\end{eqnarray}
It is easy to express the coefficients $a_{14}, b_{14}$ through the coefficient $a_{13}$,
that was determined in paper \cite{braguta}:
\begin{eqnarray}
a_{14}&=& \frac {(P^2-m_{\mu}^2)} {2 P^2} a_{13}\;, \nonumber \\
b_{14}&=& - 3   \frac {(P^2-m_{\mu}^2)} { (P^2)^2} a_{13} \nonumber
\end{eqnarray}
Second integral has the form:
\begin{eqnarray}
&\int& d \rho \frac {k_\gamma^\alpha k_\gamma^\beta k_\gamma^\sigma k_\gamma^\rho}
{(p_\mu-k_\gamma)^2-m_\mu^2}=
a_6 ( g^{\alpha \beta} g^{\sigma \rho} + g^{\alpha \sigma } g^{ \beta \rho} + g^{\alpha \rho } g^{ \beta \sigma} )+ \\ \nonumber
&&b_6 (g^{\alpha \beta} P^\sigma P^\rho + g^{\alpha \sigma} P^\beta P^\rho + g^{\alpha \rho} P^\beta P^\sigma +
g^{\beta \sigma} P^\alpha P^\rho + g^{\beta \rho} P^\alpha P^\sigma+ g^{\sigma \rho} P^\alpha P^\beta )+ \\ \nonumber
&&c_6 (g^{\alpha \beta} v^{\sigma \rho} + g^{\alpha \sigma} v^{\beta \rho} + g^{\alpha \rho} v^{\beta \sigma} +
g^{\beta \sigma} v^{\alpha \rho} + g^{\beta \rho} v^{\alpha \sigma}+ g^{\sigma \rho} v^{\alpha \beta} )+ \\ \nonumber
&&d_6 (g^{\alpha \beta} p_{\mu}^\sigma p_{\mu}^\rho + g^{\alpha \sigma} p_{\mu}^\beta p_{\mu}^\rho + g^{\alpha \rho} p_{\mu}^\beta p_{\mu}^\sigma +
g^{\beta \sigma} p_{\mu}^\alpha p_{\mu}^\rho + g^{\beta \rho} p_{\mu}^\alpha p_{\mu}^\sigma+ g^{\sigma \rho} p_{\mu}^\alpha p_{\mu}^\beta )+ \\ \nonumber
&&e_6 P^{\alpha} P^{\beta} P^{\sigma} P^{\rho} + 
f_6 (P^{\alpha} P^{\beta} P^{\sigma} p_{\mu}^{\rho} + P^{\alpha} P^{\beta} p_{\mu}^{\sigma} P^{\rho} + 
P^{\alpha} p_{\mu}^{\beta} P^{\sigma} P^{\rho} + p_{\mu}^{\alpha} P^{\beta} P^{\sigma} P^{\rho}) +\\ \nonumber
&&g_6( v^{\alpha \beta} v^{\sigma \rho} + v^{\alpha \sigma } v^{ \beta \rho} + v^{\alpha \rho } v^{ \beta \sigma} ) + \\ \nonumber
&&h_6 (p_{\mu}^{\alpha} p_{\mu}^{\beta} p_{\mu}^{\sigma} P^{\rho} + p_{\mu}^{\alpha} p_{\mu}^{\beta} P^{\sigma} p_{\mu}^{\rho} +
p_{\mu}^{\alpha} P^{\beta} p_{\mu}^{\sigma} p_{\mu}^{\rho} + P^{\alpha} p_{\mu}^{\beta} p_{\mu}^{\sigma} p_{\mu}^{\rho})
+ i_6  p_{\mu}^{\alpha} p_{\mu}^{\beta} p_{\mu}^{\sigma} p_{\mu}^{\rho},
\end{eqnarray}
where the following designation was used:~ $v^{\alpha \beta} = P^{\alpha} p_{\mu}^{\beta} + p_{\mu}^{\alpha} P^{\beta}$. 
The coefficients in this integral can be expressed through the integrals calculated in paper\cite{braguta}:
\begin{eqnarray}
i_6 &=& \frac {Pp_\mu} {p_{\mu}q}~ c_5 - \frac 3 4 \frac {P^2} {p_{\mu}q} ~ e_5 \\ \nonumber
d_6 &=& - \frac {p_\mu q} {3 P^2 }(P p_\mu ~ c_5 - p_\mu q~ i_6) \\ \nonumber
h_6 &=& \frac 1 {P^2} ( p_\mu q ~ c_5 - P p_\mu i_6) \\ \nonumber
g_6 &=&  \frac 1 {2 P^2} (p_{\mu} q~ e_5 - P p_\mu ~h_6 - d_6) \\ \nonumber
c_6 &=& - \frac 1 2 (2 P p_\mu ~ g_6 + m_{\mu}^2 ~ h_6) \\ \nonumber
a_6 &=& -(P p_\mu ~ c_6 + m_{\mu}^2 ~ d_6) \\ \nonumber
b_6 &=& - \frac 1 {Pp_\mu} (m_{\mu}^2 ~ c_6 +  \frac 1 2 ~ a_{14}) \\ \nonumber
f_6 &=& - \frac 1 {Pp_\mu} (b_6 + 2 m_{\mu}^2 ~ g_6) \\ \nonumber
e_6 &=& - \frac 1 {Pp_\mu} (m_{\mu}^2~ f_6 + \frac 1 2 ~ b_{14}) \\ \nonumber
\end{eqnarray}

\newpage
\setlength{\unitlength}{1mm}
\begin{figure}[ph]
\bf
\begin{picture}(150, 200)

\put(10,140){\epsfxsize=10cm \epsfbox{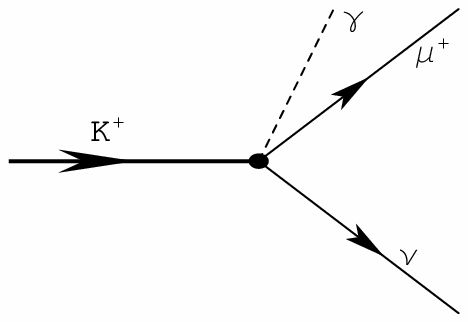}}
\put(28,140){Fig.1a}

\put(80,140){\epsfxsize=10cm \epsfbox{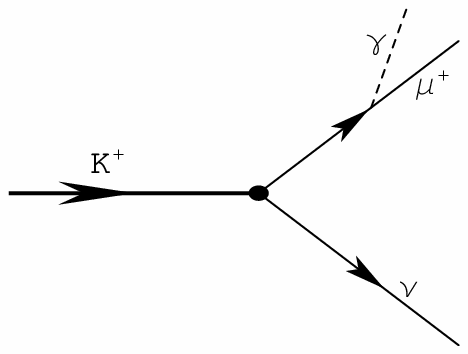}}
\put(98,140){Fig.1b}

\put(45,40){\epsfxsize=10cm \epsfbox{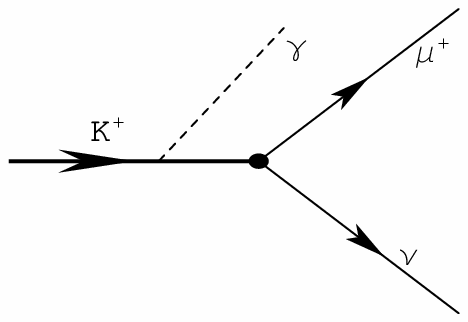}}
\put(64,40){Fig.1c}

\end{picture}
\end{figure}

\newpage
\setlength{\unitlength}{1mm}
\begin{figure}[ph]
\bf
\begin{picture}(150, 200)

\put(5,170){\epsfxsize=8cm \epsfbox{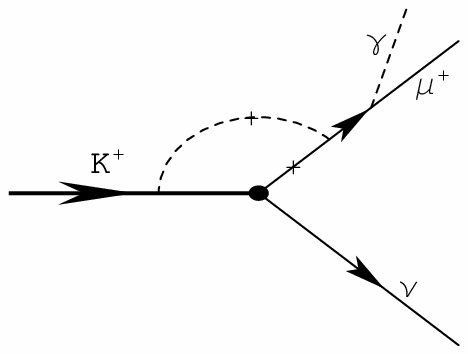}}
\put(23,175){Fig.2a}

\put(95,170){\epsfxsize=8cm \epsfbox{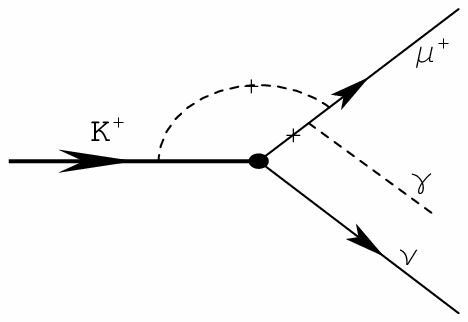}}
\put(123,175){Fig.2b}

\put(5,110){\epsfxsize=8cm \epsfbox{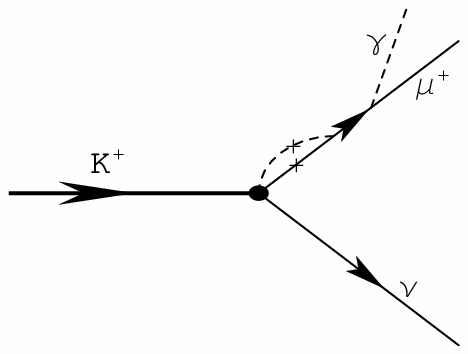}}
\put(23,115){Fig.2c}

\put(95,110){\epsfxsize=9cm \epsfbox{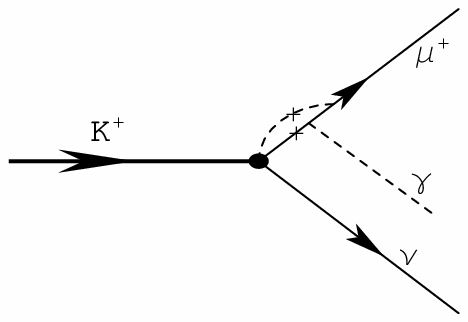}}
\put(123,115){Fig.2d}

\put(5,50){\epsfxsize=8cm \epsfbox{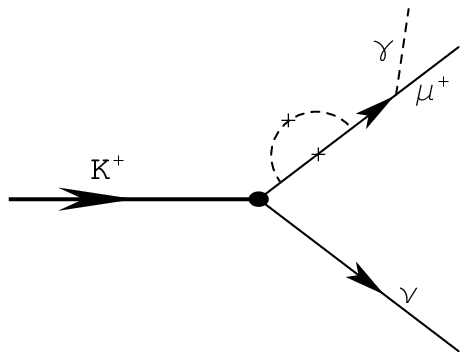}}
\put(23,55){Fig.2e}

\put(95,50){\epsfxsize=9cm \epsfbox{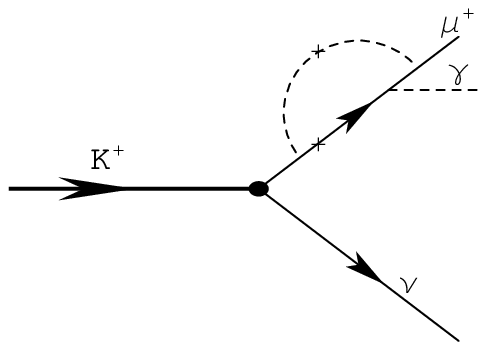}}
\put(123,55){Fig.2f}

\put(50,0){\epsfxsize=9cm \epsfbox{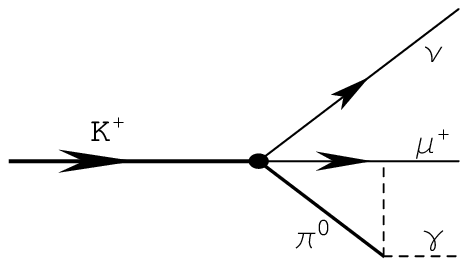}}
\put(73,5){Fig.2g}

\end{picture}
\end{figure}

\newpage

\begin{figure}[ph]
\begin{picture}(150, 200)
\put(-0,-100){\epsfxsize=17cm \epsfbox{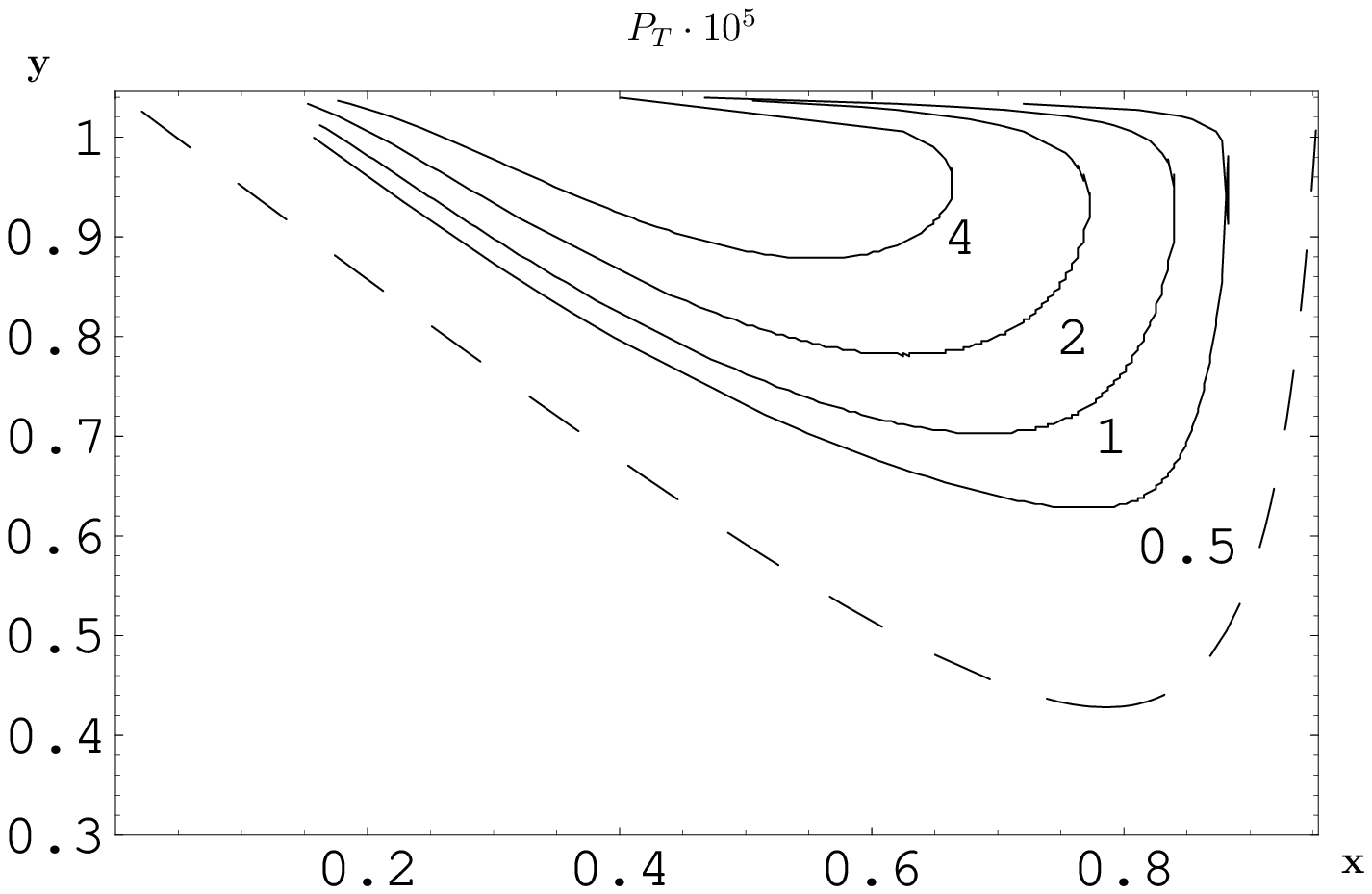}}
\put(80,40){{\bf Fig. 3}}
\end{picture}
\end{figure}

\newpage
\setlength{\unitlength}{1mm}
\begin{figure}[ph]
\bf
\begin{picture}(150, 200)
\put(40,130){\epsfxsize=9cm \epsfbox{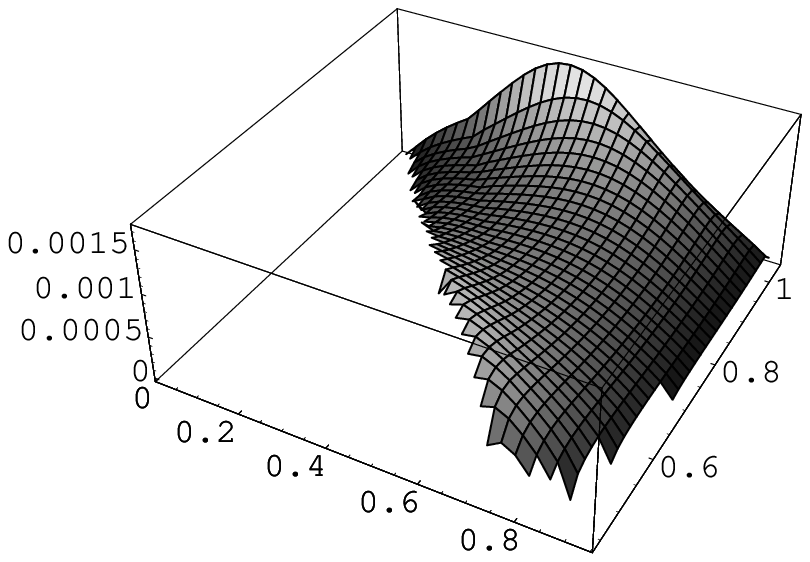}}
\put(80,120){Fig. 4}
\put(75,133){$\bf x$}
\put(125,145){$\bf y$}
\put(30,180){$\bf P_T$}

\put(40,25){\epsfxsize=10cm \epsfbox{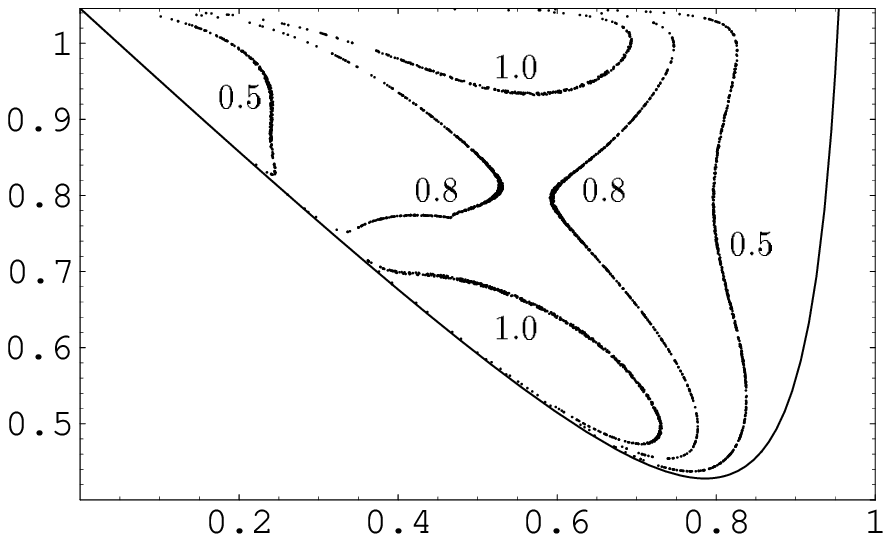}}
\put(90,90){$ P_T\cdot 10^{3}$}
\put(95,21){$\bf x$}
\put(35,60){$\bf y$}

\put(80,10){Fig. 5}

\end{picture}
\end{figure}


\vspace*{0.5cm}
\noindent

\begin{thebibliography}{29}
\bibitem{bezrukov}
F.L. Bezrukov, D.S.Gorbunov, Yu.G. Kudenko, Eur. Phys. J. C30, 487(2003).
\bibitem{braguta}
V.V. Braguta, A.E. Chalov, A.A. Likhoded, Phys.Rev.D66, 034012(2002).
\bibitem{rogalyov}
R.N.Rogalyov, Phys. Lett. b521,243(2001).
\bibitem{kudenko}
Y.G. Kudenko, Yad. Fiz. 65, 269(2002).
\bibitem{furusaka}
M.Furusaka et.al., KEK Report 99-4(1999).
\bibitem{bijnens}
J. Bijnens, G. Ecker, J. Gasser, Nucl. Phys. B396, 81(1993).
\bibitem{corr}
C.~Q.~Geng, I.~L.~Ho and T.~H.~Wu,
  Nucl.\ Phys.\ B {\bf 684}, 281 (2004)
  [arXiv:hep-ph/0306165].
\bibitem{okun}
L.B. Okun, I.B. Khriplovich, Sov. J. Nucl. Phys. 6, 821(1967).
\bibitem{rewiev}
S.~Eidelman {\it et al.}  [Particle Data Group],
  Phys.\ Lett.\ B {\bf 592}, 1 (2004).






\end{thebibliography}
\end{document}